\documentclass[useAMS,usenatbib]{mn2e}

\newcommand{\mdot}{\mbox{\em \.{M}}}

\newcommand{\vsini}{\mbox{$v_e\,\sin\,i$}}
\newcommand{\vrad}{\mbox{$v_{\mbox{rad}}$}}
\newcommand{\teff}{\mbox{$T_{\mbox{eff}}$}}
\newcommand{\prot}{\mbox{$P_{\mbox{rot}}$}}
\newcommand{\esprit}{\mbox{\small ESpRIT}}
\newcommand{\chisq}{\mbox{$\chi^{2}$}}

\newcommand{\ha}{\hbox{$\hbox{H}\alpha$}}
\newcommand{\hb}{\hbox{$\hbox{H}\beta$}}

\newcommand{\naid}{\mbox{Na~{\sc i} {\sl D}}}
\newcommand{\caii}{\mbox{Ca~{\sc ii}}}

\newcommand{\hei}{\mbox{He~{\sc i}}}

\newcommand{\kmsec}{\,\mbox{$\mbox{km}\,\mbox{s}^{-1}$}}

\newcommand{\degrees}{\mbox{$^\circ$}}

\newcommand{\rstar}{\,\mbox{$\mbox{R}_*$}}
\newcommand{\mstar}{\,\mbox{$\mbox{M}_*$}}
\newcommand{\lstar}{\,\mbox{$\mbox{L}_*$}}

\newcommand{\msun}{\,\mbox{$\mbox{M}_{\odot}$}}
\newcommand{\rasun}{\,\mbox{$\mbox{R}_{\odot}$}}
\newcommand{\lsun}{\,\mbox{$\mbox{L}_{\odot}$}}
\usepackage{epsfig}

\title[Surface magnetic fields in CV Cha and CR Cha]
{Surface magnetic fields on two accreting T Tauri stars:  CV Cha and CR Cha}
\author[G. A. J. Hussain, et al. ]
{G. A. J. Hussain$^{1}$\thanks{E-mail:
ghussain@eso.org (GAJH)}, A. Collier Cameron$^2$, M. M. Jardine$^2$, N. Dunstone$^2$, 
\newauthor J. Ramirez Velez $^3$, H.C. Stempels$^2$, J.-F. Donati$^4$, M. Semel$^3$, G. Aulanier$^3$,  T. Harries$^5$,  
\newauthor J. Bouvier$^6$, C. Dougados$^6$, J. Ferreira$^6$,  B.D. Carter$^7$, W.A. Lawson$^8$\\
$^{1}$ ESO, Karl-Schwarzschild-Strasse 2, 85748 Garching bei {M\"unchen}, Germany\\
$^{2}$ SUPA, School of Physics and Astronomy, University of St Andrews, North Haugh, Fife KY16 9SS, 
Scotland, UK \\
$^{3}$ LESIA, Observatoire de Paris, Section de Meudon, 92195 Meudon Principal Cedex, France \\
$^{4}$ LATT, CNRS-UMR 5572, Obs. Midi-Pyr\'en\'ees, 14. Av. E. Belin, F-31400 Toulouse, France\\
$^{5}$ School of Physics, University of Exeter, Stocker Road, Exeter EX4 4QL \\
$^{6}$ Laboratoire d'Astrophysique, Observatoire de Grenoble, UMR 5571, CNRS,  Univ. J. Fourier, BP 53, 39041 Grenoble Cedex 9, France \\
$^{7}$ Faculty of Sciences, University of Southern Queensland, Toowoomba 4350, Australia \\
$^{8}$ School of Physical, Environmental and Mathematical Sciences, Univ. of New South Wales,  Australian Defence Force Academy, \\
Canberra, ACT 2600, Australia 
}
\begin{document}

\date{Accepted . Received ; in original form }

\pagerange{\pageref{firstpage}--\pageref{lastpage}} \pubyear{2002}

\maketitle

\label{firstpage}

\begin{abstract}
We have produced brightness and magnetic field maps of the surfaces of CV Cha and CR Cha: two actively accreting G and K-type T Tauri stars in the Chamaeleon I star-forming cloud with ages of 3-5\,Myr. 
Our magnetic field maps show evidence for strong, complex multi-polar fields
similar to those obtained for young rapidly rotating main sequence stars. 
Brightness maps indicate the presence of dark polar caps and low latitude spots -- these brightness maps are very similar to those obtained for other pre-main sequence and
rapidly  rotating main sequence stars.

 Only two other classical T Tauri stars have been studied using similar techniques so far: V2129 Oph and BP Tau.  
CV Cha and CR Cha show magnetic field patterns that are significantly more complex than those recovered for BP Tau, a fully convective T Tauri star.   

We discuss possible reasons for this difference and suggest that the 
complexity of the stellar magnetic field is related to the convection zone; 
with more complex fields being found in T Tauri stars with 
radiative cores (V2129 Oph, CV Cha and CR Cha).  
However, it is clearly necessary to conduct magnetic field studies of T Tauri 
star systems, exploring a wide range of stellar parameters in order to 
establish how they affect magnetic field generation,  
and thus how these magnetic fields are likely to affect the evolution of T Tauri star systems as they 
approach the main sequence.

\end{abstract}


\section{Introduction}

Classical T Tauri stars  (CTTS) are pre-main sequence ($\sim$\,Myrs old) stars that are actively accreting material from their surrounding disks. As CTTS evolve to become ZAMS late-A to M-type stars, their circumstellar disks will dissipate and may ultimately condense  into planetary systems. At this young stage, CTTS display strong variability on all timescales  at optical, X-ray and EUV wavelengths. Both strong magnetic activity and variable mass accretion from  circumstellar disks onto the central stars are the key causes of this variability.

The magnetospheric accretion model successfully explains many observational characteristics of classical T Tauri stars (Ko\" nigl 1991; Camenzind 1990; Shu et al. 1994; also see review by Bouvier et al. 2006).
In this scenario the magnetosphere of a CTTS truncates the inner edge of its circumstellar disk, material is then channelled along magnetic field lines onto the stellar surface, and accretion shocks cause hot spots to form at the site of impact on the stellar surface. There is substantial observational evidence to support this theory: photometry confirming the presence of hot spots on stellar surfaces; soft X-ray spectra suggesting accretion shock temperatures and densities; NIR and FIR/mm wavelength studies which have detected disks in these systems (see review by Herbst \& Mundt 2005).

While magnetospheric accretion models can explain many of the phenomena observed in CTTS  systems, in-depth studies suggest that the relatively simple models cannot yet withstand more quantitative tests. For example, an study of IUE observations of BP Tau found a much stronger correlation between the system's accretion rate and the sizes of hot spots than that predicted by models (Ardila \& Basri 2000).
Ardila \& Basri conclude that realistic stellar magnetic fields should 
be incorporated into the magnetospheric models.

Strong magnetic fields have been directly detected on a range of CTTS using different methods. 
One approach is to measure magnetic field strengths by measuring the 
Zeeman broadening in unpolarised, magnetically sensitive lines 
(see Johns-Krull et al. 1999 and references therein). Their measurements show that CTTS have very strong kG magnetic fields,
with magnetic pressures that tend to dominate over gas pressures at their photospheres. 

Alternatively, magnetic fields can be detected by acquiring high resolution circularly polarised spectra of magnetically active stars (see Donati et al. 2008a and references). 
If a star's  circularly polarised spectra show a definite magnetic field signature, we can obtain a map of its surface magnetic field by acquiring a time-series of circularly polarised profiles covering a full rotation period. Zeeman Doppler imaging (ZDI) techniques  are then used to invert these profiles into surface magnetic field maps  (Semel 1989, Donati \& Collier Cameron 1997, Hussain et al. 2000, Donati et al. 2007, Hussain 2000). 
Donati et al. (2007, 2008b) have acquired the first detailed surface magnetic field maps of the CTTS, V2129 Oph and BP Tau. Their maps confirm the presence of kG fields measured using intensity profiles. BP Tau, a fully convective CTTS, has a predominantly dipolar field. V2129 Oph, which has  a small radiative core appears to have a more complex, predominantly octupolar  field.

Coronal extrapolations of these magnetic field maps enable us to model the coronal topology and assess the efficiency with which the star can couple with its accretion disk. Jardine et al. (2008) use these extrapolations to model the locations of open field lines (where the  stellar wind originates) and the closed field lines (which contain the X-ray emitting plasma). They investigate the effect of allowing the disk to be truncated at different radii and note the need to include more complex field topologies in magnetospheric accretion models. 

As only two CTTS systems have been mapped to date, more observations are necessary to understand how magnetic activity and accretion processes are affected by different parameters (e.g., temperature, mass, age, accretion state) in T Tauri stars. This is crucial in order to understand how and when accretion finally switches off in T Tauri stars and they become free to spin up. 

We present magnetic field maps of the surfaces of two CTTS in the Chamaeleon I molecular cloud in this paper. In Section~2 we describe the two target stars and their main properties. We provide details of the observations and data reduction procedures used in Section~3. Section~4 is an analysis of the accretion states of both stellar systems. The imaging procedure is described and maps of the surface spots and magnetic fields are presented in Sections 5 and 6 respectively. The implications of our results are discussed and our conclusions summarised  in Section~7.

\section{Chamaeleon I stars: CV Cha and CR Cha}

The Chamaeleon I (Cha I) molecular cloud is a relatively small molecular cloud 
(M$\sim$1000\msun), with a mass of roughly 120\msun\ 
concentrated in young stars 
 (see Bally et al. 2006 for more details). 
The stellar population of Cha I has a median age of $\sim 2$ Myr, within a range of $< 1$ to $> 10$ Myr suggesting prolonged or multiple star forming episodes, and is one of the
nearest star formation regions  ($d\sim 160-165$\,pc; Wichmann et al. 1998).

\subsection{CV Cha}

CV Cha (RX J1112.5-7644, LHA 332-21, SZ 42) is a  G8-type star. 
It is one of the  brightest intermediate mass stars in the Cha I molecular cloud ($V=10.8-11.0$). It also
has a relatively short rotation period  (\prot=4.4\,d) and is bright in X-rays, suggesting strong magnetic activity levels
($\log L_{\rm X}=30.1$; Bouvier et al. 1986, Bary et al. 2008). 
Moderate levels of accretion, as evidenced by strong emission in its Balmer line profiles (\ha\ EW $\sim $60--70\,\AA),
 confirm CV Cha's status as a CTTS (Covino et al. 1992; Stempels \& Piskunov 2003, Bary et al. 2008, Reipurth et al. 1996).
The most recent estimate of its temperature by spectroscopic analysis is $T_{\rm eff}=5500$\,K and $\log g=4.0$ (Stempels \& Piskunov 2003), which agrees with the G8-type classification from photometry. Photometry of CV Cha has yielded the following colours: $1.07<B-V<1.10$, $1.40<V-I<1.42$ and an extinction of $A_v=1.67$ (Gauvin \& Strom 1992).

Given these values we can estimate the evolutionary status of the system using PMS isochrones from Siess et al. (2000). As with Donati et al. (2007), we assume that CV Cha's brightest measurement ($m_V=10.8$) still represents a star that is covered by $\sim$\,20\% spot coverage. We therefore take the unspotted brightness of CV Cha to range between $V=10.6-10.8$. 
Taking CV Cha's distance to be 160\,pc (as estimated for the molecular cloud), we find that CV Cha's absolute brightness is $M_V\sim2.90$. From PMS evolutionary tracks this corresponds to an age of $5 \pm 1$\,Myr and the mass and radius parameters shown in Table\,\ref{tab:params}. 

 The varying veiling levels from 0.0--0.3 that were measured on CV Cha by Stempels \& Piskunov (2003) introduce an uncertainty into these measurements corresponding to a photospheric variability of $\Delta V\sim 0.28$. In this case the system's brightness changes to $M_V\sim3.18$ and therefore a slightly larger age$\sim 6$\,Myr but smaller mass (1.85\msun). These uncertainties are reflected in the errors quoted in Table\,\ref{tab:params}.

\begin{table*}
 \centering
  \caption{Effective temperatures, \teff, are from Stempels \& Piskunov (2003) and Schegerer et al. (2006); $P_{\rm rot}$ are from Bouvier et al. 1986. PMS evolutionary tracks from Siess et al. (2000) and photometric measurements are used to calculate the  stellar masses and radii (see Sections 2.1 \& 2.2).  $M_{\rm core}$ and $R_{\rm core}$ are the mass and radius of the radiative cores. The errors quoted here account for  changing brightness levels in these stars; other potential uncertainties, e.g., in the distances have not been accounted for. }
  \begin{tabular}{@{}lllllllll@{}}
\hline
        &     \multicolumn{2}{l}{Published parameters}   & \multicolumn{6}{l}{Values computed using PMS isochrones (Siess et al. 2000)} \\
Target  &T$_{\rm eff}$(K) & P$_{\rm rot}$(d) & \mstar (\msun) & \rstar (\rasun)  & \lstar\ (\lsun) 
& $M_{\rm core}$ (\mstar) & $R_{\rm core}$ (\rstar)& Age (Myr)   \\
CV Cha & 5500       & 4.4    &  2    $\pm 0.15$ & 2.5 $\pm  0.4$ & 7.7 $\pm 1.5$ 
      & 0.92$ \pm 0.05$        & 0.75 $\pm  0.05$          & 5.0 $\pm 1$  \\
CR Cha & 4900       & 2.3    & 1.9 $\pm 0.15$ & 2.5 $\pm 0.3$  & 3.8 $\pm 1.1$  
      & 0.65$\pm 0.06$                & 0.57 $\pm 0.1$   & 3 $\pm 1$   \\
 \hline
\end{tabular}
\label{tab:params}
\end{table*}

\subsection{CR Cha}

CR Cha (LK\ha, SZ 6)  is a K2-type star with a \teff$=4900$\,K, of a similar brightness to CV Cha ($m_V=11.2$) and a shorter rotation period (\prot=2.3\,d) (Bouvier et al. 1986, Schegerer et al. 2006). 
 Like CV Cha it is also a strong emitter in X-rays ($\log L_{\rm X}=30.17$; Feigelson et al. 1993; Robrade \& Schmitt 2006) and  displays moderate accretion levels (\ha $\sim -29.5-34$\,\AA; Reipurth et al. 1996, Guenther et al. 2007). Previous estimates of its mass, luminosity and age from D'Antona \& Mazzitelli (1994) evolutionary tracks suggest the following: \mstar$=1.2$\msun, \lstar$=3.3$\msun and $T=1$\,Myr (Natta et al. 2000).
 
As with CV Cha, by assuming an unspotted $m_V=11.0$,  correcting for an extinction of $A_v=1.37$ (Gauvin \& Strom 1992), we find $M_V= 3.63-3.86$ (depending on uncertainties in its reported distance). Fitting this to Siess et al.'s (2000) evolutionary tracks we find a higher estimated age ($3 \pm 1$\,Myr) and a higher mass than previously calculated (see Table\,\ref{tab:params}). 
Further uncertainties in the mass and age measurements could be introduced by changing levels of veiling in the system.  There are no veiling measurements of CR Cha, but as it accretes less  than CV Cha (see Section\,\ref{sec:accretion}) we assume a smaller veiling variability than that seen on CV Cha. Given veiling levels up to $\sim 0.2$, we find that CR Cha's age could vary between 3--4\,Myr and have a mass ranging between 1.75\msun\ and 1.9\msun. The corresponding uncertainties are reflected in Table\,\ref{tab:params}.
 
 It should also be noted that a source of uncertainty on ages is introduced by using different evolutionary tracks and different measurements. Using the Palla \& Stahler (1999) tracks and near infrared photometry of the same stars, Luhman (2004) find ages of approximately 1--3 Myr for our two targets. The  values reported for CR Cha's reddening appear inconsistent $A_J=0.00$ (Luhman 2004) compared to $A_V=1.37$ (Gauvin \& Strom 1992). 
On the Siess et al. (2000) tracks with these near infrared measurements, CR Cha is found to be  4.5Myr and 1.7\msun while  CV Cha is 4.3Myr and 2\msun, which is consistent with the range of values we obtain.
 
Using the Palla \& Stahler (1999) tracks and the optical photometric measurements we use here (in Table\,\ref{tab:params}), the ages of the stars are found to be similar to our measurements, with CR Cha being between 3.5-4Myr and CV Cha between 4.5-5Myr. However, the masses are found to be slightly lower (yet consistent with the uncertainties quoted in Table\,\ref{tab:params}): CR Cha is approximately 1.7\msun\ and CV Cha is about 1.8\,\msun. 
Hence the uncertainties introduced by varying levels of veiling and spot coverage are comparable or to the differences between theoretical PMS tracks.

\section[]{Observations and data reduction}

Spectro-polarimetric observations were acquired over five nights, from 2006 April 09--13, at the Anglo-Australian Observatory with an aim to map brightness and magnetic field distributions as well as monitor the accretion states of both stars. 
We mounted the visitor polarimeter, SemelPol, at the Cassegrain focus
of the 3.9-m 
 Anglo-Australian Telescope (AAT),  coupled with the UCLES spectrograph. Further details of the instrument 
configuration used can be found in Donati et al. (2003, 1999) and Donati \& Collier Cameron (1997). 

As CV Cha and CR Cha have rotation periods of 4.4\,d and 2.3\,d respectively it is necessary to obtain observations evenly spread over a period of at least five nights in order to obtain full phase coverage on both systems. 
Variable weather conditions adversely affected  the second half of the observing run so phase coverage, while in principle sufficient to cover 80\% of the surfaces of the two stars, 
remains incomplete and the S:N of the spectra variable.
 In total, we obtain 13 Stokes I and V spectra of the photosphere of CV Cha and
10 Stokes I and V spectra of the photosphere of CR Cha with sufficiently high S:N for the subsequent analysis. 
Table\,\ref{tab:obs} shows a log of the observations used in this paper.

The data were processed using \esprit, a data reduction package that optimally extracts
both polarised (Stokes V) and unpolarised (Stokes I) spectra (Donati et al. 1997) . 
The wavelength coverage ranges from 4376\AA\ to 6820\AA. 
This encompasses several accretion-sensitive diagnostics (e.g., \ha, \hei) as well as
over 3000 photospheric line profiles (some profiles are duplicated due to order overlap).

Doppler imaging and Zeeman Doppler imaging techniques ideally require 
spectra with S:N of over 1000 per resolution element to obtain reliable surface maps.
In order to boost S:N in our spectra, we use the technique of Least Squares Deconvolution (LSD; Donati et al. 1997, ). 
 This  technique is similar to a cross-correlation technique. It essentially
sums up the signal from thousands of photospheric
lines and thus fully exploits the wavelength coverage of echelle spectra.
For intensity spectra, LSD assumes that all weak to medium-strength lines have
essentially the same shape and simply scale in  depth with the line depth for each line --
the line depths are obtained from VALD 
or Kurucz model atmospheres 
(Kupka et al. 1999, Kurucz 1993, 2005)  that match the
spectral type of the star and are used to construct a line mask -- 
the observed spectra are then essentially cross-correlated with
the line mask constructed for each star.

When applying LSD to non-accreting stars, strong
 chromospheric lines are masked out when generating LSD profiles as these lines invalidate the
  assumptions used in LSD.
 Before applying LSD to these actively accreting stars, we carried out careful checks to ensure
 that strong accretion-sensitive diagnostics were effectively masked out. 
 The line mask constructed by Stempels \& Piskunov (2003) 
 selects absorption lines that are unaffected by emission. We use this to construct a line-list to use for LSD.  A separate line-list was created using the VALD database.
  A comparison of the LSD profiles obtained with the LSD line-list from  Stempels 
 \& Piskunov (2003) and the LSD line-list from VALD  shows no significant difference in the shape of the
 LSD profiles obtained.  
 
  We reject any input spectra with peak S:N levels of below 30 in our subsequent analysis. 
 Spectra with input peak S:N$<30$ result in  LSD profiles with shapes and line depths that are inconsistent with the LSD profiles for spectra with higher S:N. The reason for this is likely due to incorrect continuum normalisation at low S:N levels.
 We are confident that the LSD profiles we use here are free of potential contaminants due to  line profiles that are sensitive to accretion. The LSD profiles used for the CV Cha imaging analysis 
 presented in this paper were  produced using the line-lists derived from the earlier study of CV Cha by Stempels \& Piskunov (2003). The ones for CR Cha were constructed using the VALD database, but omitting the emission-affected regions marked out in the CV Cha line-list.

\begin{table*}
 \centering
  \caption{Table of observations. Longitude is the central longitude at that particular rotation phase: $l=(1-\phi)*360$\degrees. Phases that were excluded from the final image reconstructions are marked as ``excluded''.}
  \begin{tabular}{@{}lllllll@{}}
  \hline
   UT Date         & MJD        	&	Target      &Phase	& Longitude (\degrees)          & S:N (inp/LSD)  & Comment\\
   \hline
2006 Apr 09    &  53834.4322 	& 	CV Cha	&0.000	& 0.0/360		&60/2280 & excluded: bootstrap analysis\\
			& 53834.4797 	& 	CR Cha	&0.000	& 0.0/360		&47/1850&\\
			 & 53834.5267 	& 	CV Cha    &0.022	& 352	&51/2040 &\\
                         &  53834.5734  & 	CR Cha	&0.042  	& 345	&60/2350&\\
                         & 53834.6216 	& 	CV Cha    &0.044    	& 344	&52/2190 &\\
                         & 53834.6680   & 	CR Cha	&0.083     	& 330	&50/1768&\\
                         & 53834.7203 	& 	CV Cha    &0.066	& 336	&50/1885& excluded: bootstrap analysis\\
2006 Apr 10     & 53835.3886	& 	CV Cha    &0.218	& 282	&54/2230& excluded: bootstrap analysis\\
                          & 53835.4352  & 	CR Cha	&0.417	& 210	&55/2030&\\
                          & 53835.4830 	& 	CV Cha    &0.239	& 274	&44/1260&\\
                          & 53835.5290	& 	CR Cha	&0.457	& 195	&37/910&\\ 
                          & 53835.5754 	& 	CV Cha    &0.260	& 266	&38/1280&\\
                          & 53835.6756 	&	CV Cha    &0.283	& 258	&47/1550&\\
                          & 53835.7219 & 	CR Cha	&0.541	& 165	&33/1112&\\
2006 Apr 12     & 53837.3794 	& 	CV Cha    &0.670	& 119	&68/2370&\\
			& 53837.4254  & 	CR Cha	&1.282	& 258	&63/2240&\\
                          & 53837.4717 	& 	CV Cha    &0.691	& 111	&72/2760&\\
                          & 53837.5178 & 	CR Cha	&1.322	& 244	&50/1000&\\
2006 Apr 13     & 53838.5409 	& 	CV Cha    &0.934	& 24		&50/1560&\\
			& 53838.5870   & 	CR Cha	&1.787	& 77		&47/1716&excluded: change in accretion state\\
                          & 53838.6333 	& 	CV Cha    &0.955	& 16		&58/2220&\\
                          & 53838.6803	& 	CR Cha	&1.827	& 62		&50/1859&excluded: change in accretion state\\
                          & 53838.7265 	& 	CV Cha    &0.977	& 8		&47/1780&\\
 \hline
\end{tabular}
\label{tab:obs}
\end{table*}

\section[]{Accretion diagnostics}
\label{sec:accretion}

\begin{figure*}
\epsfig{file=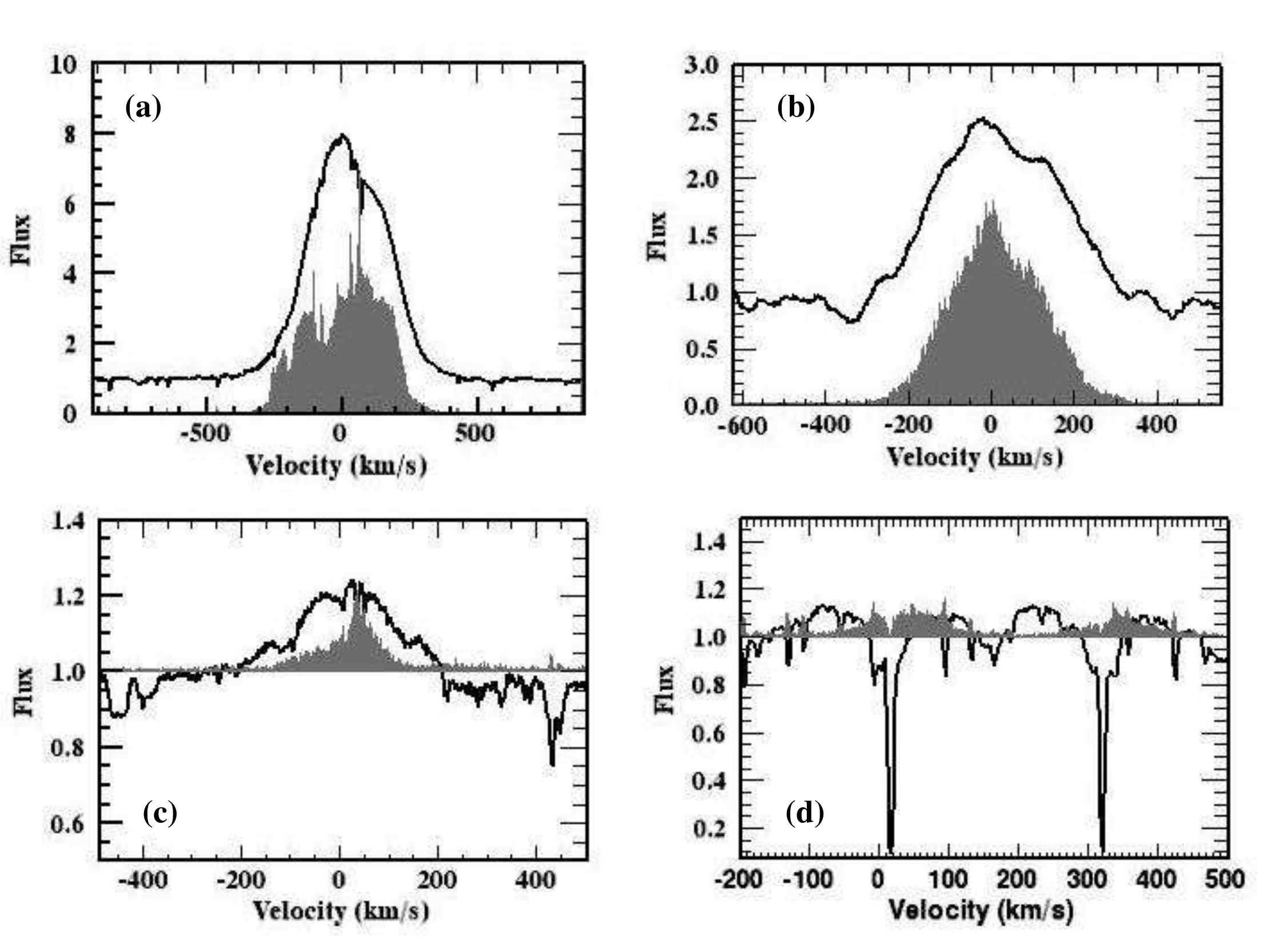, width=14cm}
\caption{Accretion diagnostics for CV Cha: 
The solid lines are the mean line profiles for each diagnostic and the grey solid plot 
represents the variance profile (multiplied by a factor of 10 for clarity). 
(a) The \ha\ profile and (b) the \hb\ profiles are clearly in emission and relatively
symmetric in shape;  (c) the \hei\ profile is in emission and (d) the \naid\ profile shows
two narrow absorption features caused by interstellar absorption. The variance profiles for \ha\ and \hei\ show a strong changing
contribution to the red wing of the line profiles, indicating inflows onto the star.
}
\label{fig:emit_cvcha}
\end{figure*}

Several strong accretion-sensitive diagnostics can be used
to characterise the accretion states of both stars at this epoch.
An analysis of the \ha, \hb, \hei, and \naid\ line profiles  shows that both stars 
are actively accreting at this epoch. 
  No polarisation was detected in the Stokes V  diagnostics of these features, presumably
 due to insufficient S:N.
 
\begin{figure*}
\epsfig{file=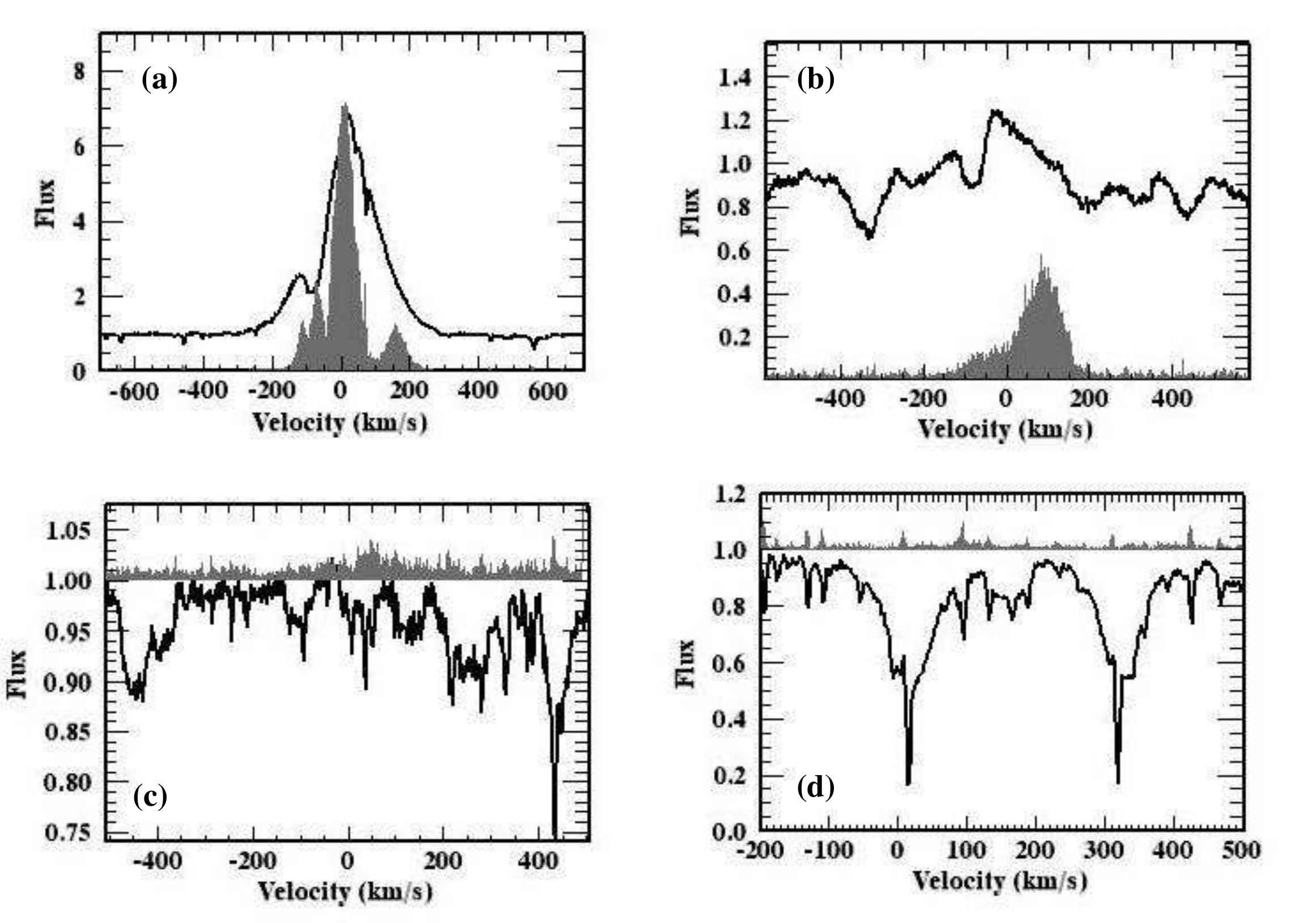, width=14cm}
\caption{Accretion diagnostics for CR Cha: 
As in Fig.~\ref{fig:emit_cvcha}, the solid lines represent the mean profiles and the grey solid 
plots are the variance profiles for each emission line: (a) \ha, (b) \hb, (c) \hei, (d) \naid.
The \hb\ profile shows a strong red-shifted component in both the mean and variance profiles, indicating
inflows. The \ha\ profile also has a redshifted variance component as well as a strong blue-shifted contribution to the mean line profile indicating stellar winds.
CR Cha is clearly not accreting as strongly as CV Cha as the \hei\ diagnostic is in absorption.
}
\label{fig:emit_crcha}
\end{figure*}

Figs.~\ref{fig:emit_cvcha} and \ref{fig:emit_crcha} show the mean and variance line profiles
for these diagnostics in CV Cha and CR Cha respectively.
In these figures, the mean line profiles are the solid black lines and the variance profiles are 
represented by grey filled in plots. 
 Variances have been scaled by a factor of 10 (and offset by 1.0 for \hei\ and \naid) so that they can be plotted on the same scale as the mean line profiles.
\ha\ is strongly in emission for both stars confirming their previous classifications as classical T Tauri stars. We measure their equivalent widths using the dedicated program in the spectral analysis
package DIPSO (Howarth, Murray \& Mills 1994).
The two stars' mean \ha\ equivalent widths are $-51$\,\AA\ and $-25$\,\AA\ for  CV Cha and CR Cha respectively; these values are between 5-10\AA\ lower than previous measurements for both stars (Guenther et al. 2007;  Reipurth et al. 1996).

While the mean \ha\ profile for CV Cha is almost symmetric, CR Cha's \ha\ profile possesses a significant blue-shifted component, indicating outflows and winds on this system. The \ha\ variance profiles in both stars show variations that can be separated into three components: the blue-shifted component indicates varying winds; the central component is likely due to varying accretion onto the star; and the red-shifted component indicating inflows onto the star.  
Further evidence of inflows in CR Cha can be found in its H$\beta$ profile: the mean line profile has a strong asymmetry in its red wing and the greatest variations are also found in the red-shifted part of the profile also supporting the presence of inflows. 

The \hei\ line profile in CR Cha is in absorption, while in CV Cha it is clearly in emission: further evidence that  CV Cha is accreting more strongly.
The \naid\ profiles in both stars show evidence of interstellar absorption. 

\subsection{Variability analysis}  
The individual \ha\ and \hb\ profiles were examined for both stars to evaluate how
much their accretion states varied from  night-to-night (see Figs.\,\ref{fig:emit_cvcha} \& \ref{fig:emit_crcha}). 
This is also necessary to establish how stable the surface active regions were over 
the time-span of observations and therefore how reliable the maps are likely to be.
We find that CV Cha is significantly variable from night to night and therefore employ a bootstrap analysis to get the final images of this star (see Section~\ref{sec:imaging}). 
There is significant blue-shifted emission (likely due to winds) at phases  0.21--0.23, while excess emission and red-shifted emission (due to accretion inflows and hotspots)
 are detected at phases 0.95--0.06.

CR Cha's Balmer line profiles show less variation from night to night --
 until the last night (2006 April 13), when their peak strengths increase by 30\% compared to previous exposures. In our image reconstructions of CR Cha we examine the effect of omitting the 
last night's exposures on the brightness and magnetic field maps.
Both inflows and outflows (as indicated by excess red-shifted and blue-shifted emission) are indicated at phases 0.0-0.040. The last two exposures at the end of the run show how much the star's accretion state increased. These two exposures, corresponding to phases 0.78 and 0.83 show the strongest emission in both \ha\ and \hb\ line profiles.

\subsection{Estimating \mdot}  
 We can estimate the relative levels of accretion in both stars using the width of the \ha\ profile at 10\% of
its peak flux, and the  empirical relation first advanced by Natta et al. (2004). Herczeg \& Hillenbrand (2008) find that the \mdot\ measured using this method is comparable with that found using \ha\ line profile modelling, with a scatter in \mdot\ of 0.86\,dex. The mean \ha\ 10\% line widths we find for CV Cha and CR Cha are  550 and 390\,\kmsec\ respectively. This corresponds to 
 $\log\mdot_{\rm acc}$ values of -7.5 and -9.0 (\msun/yr).  Given the uncertainty of 0.86 dex, CV Cha is likely a stronger accretor than CR Cha, but substantial uncertainty remains. The variability in the \ha\ 10\% line widths was $\pm 35$\,\kmsec\ for CV Cha and $\pm 15$\,\kmsec\ for CR Cha.
 
\begin{figure*}
\epsfig{file=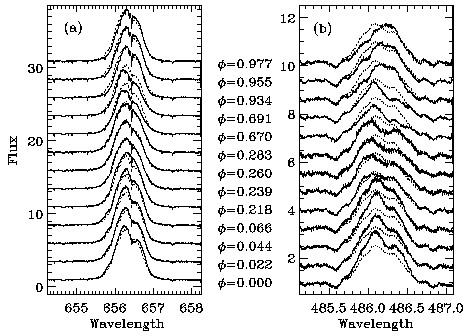, width=14cm}
\caption{Balmer line profiles for CV Cha. (a) \ha\ line profiles and (b) \hb\ line profiles.
The dotted line shows the over-plotted mean line profile, to illustrate the line variability from phase to phase more clearly.
 }
\label{fig:emit_cvchaphase}
\end{figure*}

\begin{figure*}
\epsfig{file=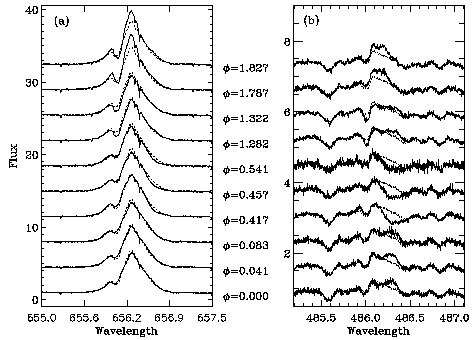, width=14cm}
\caption{As with Fig.\,\ref{fig:emit_cvchaphase} for CR Cha. The last two observations illustrate that the accretion phase of the system increased
significantly on the last night.}
\label{fig:emit_crchaphase}
\end{figure*}

\begin{figure*}
\includegraphics[width=8.5cm]{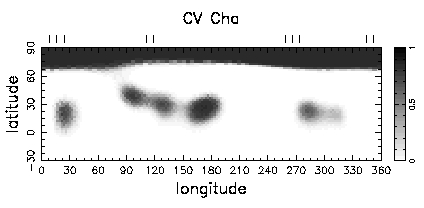}
\includegraphics[width=8.5cm]{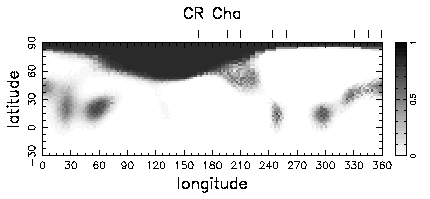}
\caption{Brightness map for CV Cha (left) and CR Cha (right). The tick marks show the phases of observation. The spot groups here have been modelled as dark cool spots.}
\label{fig:intensity}
\end{figure*}

\section{Brightness imaging}
\label{sec:imaging}
Prior to Doppler imaging, we scale all of our LSD intensity profiles to the same EW in order to  
remove the effects of changing veiling levels from all of the profiles. 
Veiling correlates poorly with polarisation and profile
distortions (Valenti, Johns-Krull \& Hatzes 2003, Donati et al. 2007, 2008b). 
In the absence of a consistent model to describe veiling, 
 it becomes necessary to remove any possible  effects due to veiling from the spectra.
 Furthermore,  veiling on CV Cha was found to be low, varying between 0.0 and 0.3
(Stempels \&  Piskunov 2003) so the effect is expected to be small. There are no published veiling measurements for CR Cha. 

We model the distortions in the LSD profiles assuming that they are caused by dark cool spots. 
It should be noted that there is some ambiguity here:  the line deformation  in de-veiled profiles is essentially the same, whether the cause is a hotspot or
a completely  dark spot with a similar area (Unruh, Collier Cameron \& Guenther 1998, Unruh et al. 2004). This assumption has also previously found to be reasonable
 for V2129 Oph (Donati et al. 2007, Shevchenko \& Herbst 1998)
and BP Tau (Donati et al. 2008b). As we can obtain good fits to the spectra using this model 
we also assume that the main causes of distortions and line asymmetries in these LSD profiles 
 are dark spots crossing the projected stellar disk as it rotates. 

In order to obtain reliable brightness and magnetic field maps we first need to 
derive accurate system parameters: \vrad, \vsini, EW, and inclination angle, $i$.
The procedure involves reconstructing a series of images using grids of different stellar parameters:
 the optimum parameters are determined by reducing $\chisq$ in the brightness imaging code 
 (Table\,\ref{tab:param_image}).
 The $\chisq$ minimisation procedure is identical to that used in previous studies (e.g., Barnes et al. 1998).
 These parameters are then used to reconstruct surface brightness and magnetic field maps.

\vrad\ is determined first as it is independent of the other parameters. 
 There is  a well-documented dependence between \vsini\ and EW when evaluated in this manner; e.g., too low a line EW and too high a \vsini\ value produce a deficit in the flux at the centre of the line profile but an excess in the wings (Collier Cameron \& Unruh 1994). 
EW and \vsini\ are therefore optimised together.  
 Note that, while EW can be measured independently the \vsini-EW
dependence needs to be minimised carefully  as together they affect the
shape of the model line profile. An error in EW  of just a few \% 
will introduce artefacts into the resulting Doppler images (see Collier
Cameron \& Unruh for further details.
As the EW has been renormalised, the optimum \vsini\ we obtain may not be the ``true'' \vsini\ value. However, it is instructive to compare this with published measurements to evaluate the effect of removing the veiling contribution from the spectra.  While our  \vsini\ value of $25 \pm 1$\kmsec\ is a little lower than the 28\kmsec\ reported by Stempels \& Piskunov (2003), it is within their measurement errors.

Once the \vrad, \vsini\ and EW values are measured $i$ and \prot\ can be estimated
in the same way, although as the images are not very sensitive to these parameters, there are much larger uncertainties associated with these two parameters. 
In the case of CR Cha, $\chisq$ minimises at 2.3\,days (i.e. the published \prot\ value).
For CV Cha, the optimum \prot\ values agrees with the published values with an uncertainty of $\pm 0.2$\,days, consistent with the error bars on the measurement.

\begin{table}
 \caption{Stellar parameters derived from parameter optimisation during Doppler imaging.}
 \begin{tabular}{@{}llll@{}}
 \hline
 Target   &   v$_{\rm rad}$ (\kmsec) & \vsini (\kmsec) & Inclination (\degrees) \\ 
  CV Cha &  22.7$\pm 1$ & 25$\pm$ 1  & 35 $\pm 10$ \\
  CR Cha &  24.0$\pm 1$ & 35$\pm$ 1 & 50 $\pm 10$ \\
 \hline
\end{tabular}
\label{tab:param_image}
\end{table}

Brightness maps are first obtained using the entire dataset: we then explore the effect on the images of omitting data from separate epochs. 
As stated earlier, there is clear evidence from the \ha\ and \hb\ profiles that the accretion state of both CV Cha and CR Cha appear to change. 

In order to assess the reliability of these maps and to exclude the most strongly accreting phases (which may introduce artefacts) we employ a bootstrap analysis for CV Cha.
We omit three out of the 13 spectra and run the imaging code, measuring the minimum \chisq\ achieved after 40 iterations for each possible combination of spectra. We then adopt the combination(s) of spectra that yield the minimum \chisq\ after this fixed number of iterations (in a similar procedure to that used for optimising the system parameters).
This approach allows us to improve on the minimised \chisq\ by a factor of two by leaving out three spectra: spectra 4, 5, 13 -- these are marked as excluded in Table~\ref{tab:obs}. The images shown in Figs.~\ref{fig:intensity} and \ref{fig:cvcha_mag} show the maps obtained from this cropped ten-spectra dataset. The locations, numbers and sizes of spots reconstructed in  spot maps from the cropped  and uncropped thirteen-spectra dataset are very similar.
The reduced \chisq, $chi^2_r$, in the map shown here is 2.5;  where the number of degrees of freedom is 240 for the CV Cha imaging dataset.
The only significant advantages to using the cropped dataset are that  the resulting images provide a better fit to the data ($\chi^2_r$=2.5 compared to 3.0), and a small noisy spot feature near 320\degrees\ longitude is removed. The overall stability of the spots reconstructed in these maps  implies that these spot groups are stable and not caused by noise artefacts. We also compare the brightness map obtained at $\chi^2_r =2.5$ with maps obtained at $\chi^2_r=5$  to check the robustness of our maps and ensure that the maps are not fitting noise. We find that the spot map presented here  is essentially unchanged compared to the $\chi^2=5$ map: this is a  further indication that the spots we reconstruct are genuinely in the data and not artefacts.

As discussed in the previous section, CR Cha's accretion state appears to elevate significantly in the last night, which may affect its surface map. 
When comparing brightness maps obtained with the full dataset and the cropped dataset (omitting the last night's spectra) we find the spot patterns 
agree well but that we can achieve lower \chisq\ values with the cropped dataset. The map we show here was derived using the cropped dataset and the 
phases excluded are marked in Table~\ref{tab:obs}. All of the spot groups reconstructed for CR Cha are robust, though the spot group between 
330\degrees--360\degrees\ longitude appears poorly constrained. The reliability of these spots was also tested further by comparing the brightness 
map obtained at $\chi^2_r =2.5$ with maps obtained at $\chi^2_r=5$ (as was done for CV Cha)  we find that all of the spot groups appear in the less 
well defined spot map; the only difference is that the feature near 330-360\degrees\ appears less well constrained at $\chi^2_r=5$; the number of degrees
of freedom in the CR Cha imaging datasets is 248.
 
We look for correlations between the \ha\ 10\% width measurements, the LSD line equivalent widths and the measured spot coverage at each phase of observation to assess whether or not these spots are hot  or cold. If the spots reconstructed are hot then one would expect an anti-correlation between the line EW and the spot coverage (as the more veiled the star would be the higher the spot coverage and the weaker the line equivalent width). We find no evidence for correlations between the LSD line equivalent width, 
the spot coverage in these maps and the \ha\ 10\% width values, supporting the assumption that the spots reconstructed here are indeed cool spots.   

The fits and residuals corresponding to CV Cha and CR Cha maps are shown in the Appendix (Fig.~\ref{fig:intfits}). The brightness maps shown in Fig.~\ref{fig:intensity} produce fits to a reduced \chisq\ level, $\chi^2_r= 2.5$. CV Cha  has a large polar cap extending down below 70\degrees\ latitude and five spots at the mid-to-low latitude regions. CR Cha shows a similar spot pattern to CV Cha, although its polar cap is asymmetric. 
The spot patterns we find here appear typical for other classical T Tauri stars, which typically show large polar caps and spots at mid to low latitudes. Recent examples of other stars with similar spot patterns are the pre-main sequence stars SU Aur (Unruh et al. 2004), HD155555 (Dunstone et al. 2008, Hatzes 1999),
 TWA 6 (Skelly et al. 2008). ZAMS G and K-type stars also show similar spot patterns, e.g., the $\alpha Per$ G-type dwarf stars: He 520, He 699 (Barnes et al. 1998), and AB Dor (K2V; Collier Cameron \& Unruh 1994).

\section{Surface magnetic field maps}

\begin{figure*}
\includegraphics[trim= 0mm 100mm 0mm 20mm]{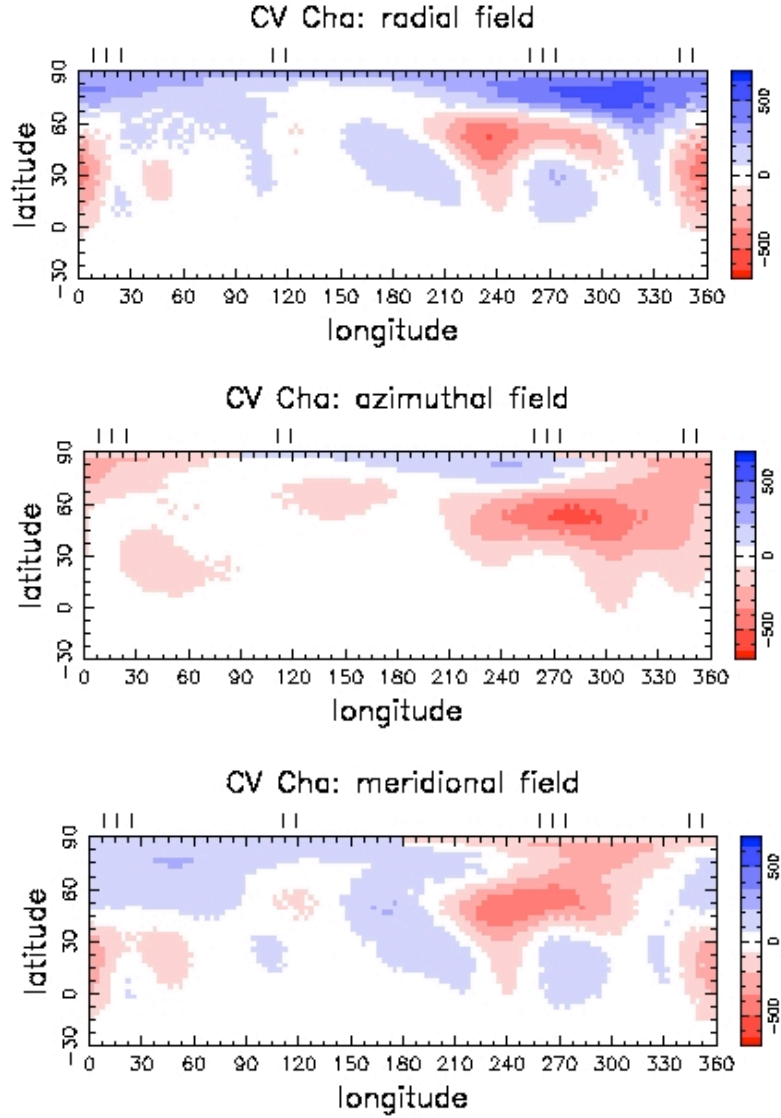}
\caption{Magnetic field maps for CV Cha: these maps fit the data to  a $\chisq_r=1.1$.
}
\label{fig:cvcha_mag}
\end{figure*}

\begin{figure*}
\includegraphics[trim = 0mm 100mm 0mm 20mm]{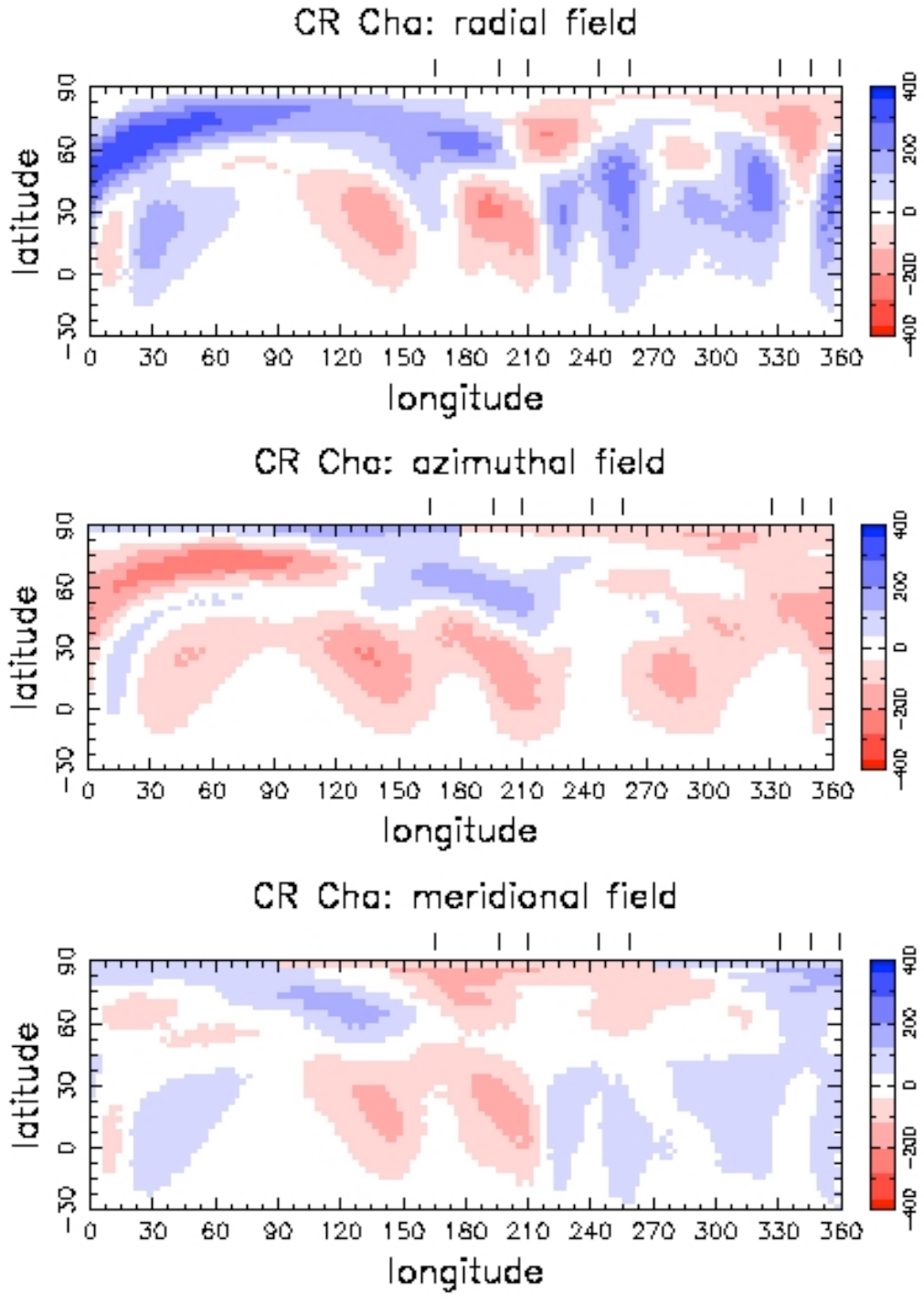}
\caption{Magnetic field maps for CR Cha: these maps fit the data to a $\chisq_r=1.0$.
}
\label{fig:crcha_mag}
\end{figure*}

We obtain magnetic field maps using LSD Stokes V profiles derived using the same intrinsic line EW as the intensity line profiles used to derive the brightness maps. The ZDI code used is described by 
Hussain et al. (2000). 
The strongest magnetic field detections were obtained for CV Cha (the brighter of the two stars) 
with a false alarm probabiliy, FAP$\sim10^{-5}$; for CR Cha only marginal detections were obtained with a FAP$\sim10^{-3}$. 
 The FAP is defined as in Donati et al. 1997 and is based on the reduced \chisq\  statistics computed both inside and outside the spectral lines in both the LSD Stokes V profile and the null polarisation profile (see Donati et al. 1997 for more details). 
However, despite this,  we find that we can recover robust magnetic field maps (Figs.~\ref{fig:cvcha_mag} \& 
\ref{fig:crcha_mag} and Figs.~\ref{fig:magfits}). 

The magnetic field maps obtained here have been derived solely from photospheric line profiles. Donati et al. (2007, 2008b) studied the other two  classical T Tauri stars, V2129 Oph, and BP Tau. 
Their maps are different as they include circularly polarised profile from photospheric lines and 
accretion-sensitive lines that originate at the base of the accretion funnels; the field of the accreting lines can be mapped using circularly polarised (Stokes V)   \caii\ line profiles. Unfortunately the wavelength range of our data does not encompass this diagnostic, and no Stokes V signatures are detected in the other accretion-sensitive diagnostics in our data.

The effect of including the accretion-sensitive diagnostics was explored by Donati et al. (2007) when producing surface magnetic field maps of V2129 Oph. They find that maps obtained from photospheric line profiles alone do accurately reflect the field topology of the non-accreting surface, even if  the strong field from the footpoints of accretion funnels cannot be recovered (see Fig.\,12 of Donati et al. 2007). 
Hence the complex topologies we recover can be compared reliably with those obtained for other pre-main sequence and main sequence stars.

The magnetic field maps for CV Cha (Fig.~\ref{fig:cvcha_mag})  show the presence of strong magnetic fields (up to 600\,G) covering most of the observable surface. The radial, azimuthal and meridional field maps trace a multipolar field. The field between longitudes 0--90\degrees\ appears significantly weaker than that on the opposite hemisphere. This may be a consequence of poor phase coverage or it may be caused by the presence of hot accretion spots between 0--120\degrees\ that essentially neutralise the magnetic field signatures from this region of the star\footnote{ The maps of V2129 Oph do show a large accretion spot with a radius of over 15\degrees\ in latitude (Fig.11 of Donati et al. 2007), in contrast to the smaller filling factors suggested by Valenti et al (1993).}.
The strongest indications of accretion from the Balmer line profiles
are near phase 0, which corresponds well with the longitudes at which the magnetic field maps appear to be affected by accretion hotspots. A faint magnetic field feature near 30\degrees\ longitude appears to be co-spatial with the dark cool spot in the brightness map -- this may mark the point of impact of an accretion funnel.
 
The magnetic field maps for CR Cha (Fig.~\ref{fig:crcha_mag}) also show strong complex multipolar magnetic field patterns covering the entire observable star. As this star has a higher \vsini\ value, its maps potentially have better spatial resolution than CV Cha. Unlike CV Cha, CR Cha's surface seems to be relatively unaffected by accretion. The maximum field obtained is $\pm 400$\,G for these maps, i.e., somewhat suppressed compared to CV Cha. This is likely due to CV Cha's being slightly fainter and its circularly polarised profiles being noisier than those of CV Cha.
 CV Cha and CR Cha are found to have  average magnetic filling factors of 0.30 and 0.10 respectively.
Note, however,  that these are probably lower in CR Cha due to the lower
input S:N of the data (as opposed to reflecting a genuine weaker magnetic field).

\section{Discussion}

\subsection{Surface magnetic field maps of T Tauri stars}

Two other CTTS have been imaged using similar Doppler imaging techniques: V2129 Oph and BP Tau (Donati et al. 2007, 2008b). Both of these are younger and less massive than CV Cha and CR Cha: 
BP Tau (M=0.70\msun, R=1.95\rasun, \teff=4055\,K, \prot=7.6\,d, Age$\sim$1.5\,Myr) 
is fully convective, and 
V2129 Oph (M=1.35\msun, R=2.4\rasun, \teff=4200\,K, \prot=6.53\,d, \vsini=15\,\kmsec, Age=2\,Myr) has a small radiative core 
(M$_{\rm core} \sim $0.1\mstar\ and R$_{\rm core} \sim$0.2\rstar, Age$\sim$2\,Myr). 

BP Tau shows a simple magnetic field structure, consisting predominantly of a dipole and octupole
field component of similar strengths; in particular its azimuthal component is relatively weak. 
V2129 Oph, on the other hand, shows a more complex field structure, dominated by an octupole and a much weaker dipole: its azimuthal field component is relatively strong (of comparable strength to the radial field map) in the maps recovered using only photospheric line profiles). The surface magnetic field maps of V2129 Oph obtained using photospheric line profiles only show an azimuthal field strength that is comparable to the radial field. Adding the circularly polarised profiles from the accretion line profiles strengthens the dipole component of the global field but does not alter the topology of the azimuthal field significantly. 

If we compare our maps to the V2129 Oph brightness maps and the magnetic field maps 
obtained using only the LSD photospheric line profiles (i.e. right hand column of Fig.\,12 in Donati et al. 2007), we are comparing maps that have been reconstructed with essentially  the same technique. 
As CV Cha and CR Cha have \vsini\ values that are almost double that of V2129 Oph, their maps could potentially have twice the spatial resolution: the maximum resolution for CV Cha and CR Cha corresponds to $\sim$ 11\degrees\ and 8\degrees\ latitude respectively, while for V2129 Oph it is closer to 20\degrees. However, the poor weather and resulting poor phase coverage actually reduces the maximum spatial resolution of our maps, degrading the CV Cha maps to a similar spatial resolution as V2129 Oph (i.e. $\sim$20\degrees\ latitude). CR Cha maps have slightly better resolution, as evidenced by the smaller spots reconstructed in the brightness maps, but still a factor of two worse than the maximum achievable spatial resolution.

A comparison of all three CTTS brightness and magnetic field maps reveals some similarities: all brightness maps show a polar cap with well defined spots at lower, near-equatorial latitudes with radial and azimuthal magnetic field components being of approximately equal strength. 

The magnetic field maps also appear similar: V2129 Oph, CV Cha and CR Cha all contain strong azimuthal field
of approximately equal strength to the radial field in each map. 

\subsubsection{Spherical harmonic decomposition of radial  fields}
\begin{figure*}
\includegraphics[width=18cm]{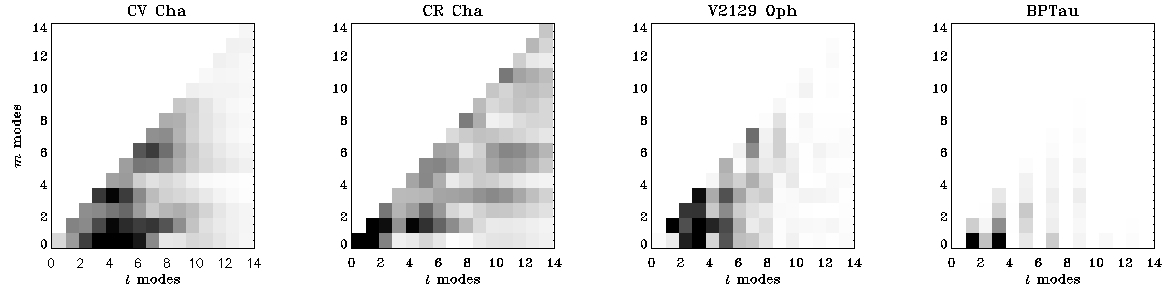}
\caption{Dominant $l,m$ modes for the radial magnetic field maps of
 CV Cha, CR Cha, V2129 Oph (Donati et al. 2007) and BP Tau (Donati et al. 2008b).
All of the maps used for this analysis were obtained using photospheric line profiles only. }
\label{fig:modes}
\end{figure*}

A more detailed analysis of these maps can be conducted by decomposing the surface magnetic field maps into their spherical harmonic components. Fig.\,\ref{fig:modes} shows the  dominant $l,m$ modes in the radial magnetic field maps 
 of the three radiative CTTS: CV Cha, CR Cha, V2129 Oph, as well as the fully convective CTTS, BP Tau.
 These $l,m$ modes are essentially a measure of the complexity of the magnetic field distributions in each of these stars. An octupole is at $l=3$ and the spatial resolution of these maps should correspond to approximately $l=10$.
 The greyscale in each plot has been  renormalised so the strongest mode in each plot is black and the weakest modes are white (so the relative strengths of the modes in each star are comparable).  In general CV Cha, CR Cha and V2129 Oph show
 more complex fields than those reconstructed in the fully convective BP Tau, 
 which has power concentrated in very low $l,m$ modes.  BP Tau has its highest
 power concentrated in low modes $l=1$, $l=3$ and $m=0$, while the other CTTS have
 power in higher $l,m$ modes.

 We can quantify the statistical significance of the $l,m$ modes (and thus
the complexity of the field) as follows:
we produce a series of images omitting the strongest $l,m$ mode combinations and then 
measure the level of fit with the original dataset. 
If the change in $\chisq \ge 3$ then it is judged to be a mode 
with significant power.
 The highest modes with power in the CV Cha map at ($l=4,m=3$) and ($l=7,m=6$) have significance levels $\Delta\chi^2>4$s; while those for the CR Cha map at ($l=8,m=8$), ($l=10,m=3$) and ($l=11,m=11$) all have $\Delta\chi^2\sim3$.

There is a possibility that the differences between the CV Cha and
CR Cha images compared to the V2129 Oph and BP Tau images are solely
caused by differences in the stellar \vsini. 
We investigate that suggestion here.
Higher \vsini\ values enable higher spatial resolution in Doppler images 
and therefore higher $l,m$ modes to be detected in the surface maps:
the optimal spatial resolution attainable with this instrumental setup 
for CR Cha is 8\degrees\ in latitude, 
for CV Cha this is 12\degrees\ and for BP Tau it is approximately 27\degrees.
In practise, the actual spatial 
resolution achieved in the images presented here 
is considerably worse  due to the poor phase coverage.

 In order to investigate the significance of the complexity in our maps 
compared to that obtained for BP Tau and V2129 Oph we carry out the following test. We simulate data for BP Tau and V2129 Oph using the magnetic field maps 
obtained for CR Cha, 
incorporating the stellar parameters, phase coverage and S:N levels obtained by Donati et al. (2007, 2008b). 
Images are then reconstructed from these simulated spectra.
These images are deconvolved into their spherical harmonic components
and we then employ the procedure described above to determine the 
significance of the most complex modes observed in these simulated images
(i.e. by producing a series of images with the most complex modes
removed and recomputing the fit to the data).
 
We find that the simulated V2129 Oph maps show a similar level of 
complexity as the V2129 Oph maps reconstructed by Donati et al. (2007).
In contrast, the simulated BP Tau map is significantly more complex than that 
reconstructed by Donati et al. (2008b).
The simulated BP Tau mode map shows significant power at ($l=1, m=1$)
 with a $\Delta\chi^2> 100$, and 
the more complex modes, e.g., at ($l=5, m=5$) has a significance level of 
$\Delta\chi^2=3$. 
There is no corresponding mode in the observed BP Tau map.
Hence even though BP Tau has a lower \vsini\ 
using this technique we can still 
reconstruct significant power at $l=5,m=5$ using 
data with the same S:N and phase coverage as that used by
Donati et al. (2008b). We conclude therefore that the magnetic
field on BP Tau is indeed less complex than that obtained
for V2129 Oph, CV Cha and CR Cha.

\subsection{What affects the complexity of the magnetic field in T Tauri stars?}

 We consider whether the stellar rotation period is the main 
cause of differences between the radiative CTTS and BP Tau.
However, as CR Cha rotates almost twice as fast as CV Cha and 
V2129 Oph, yet shows a similar level of complexity, this does not appear
to affect the complexity of the magnetic field in these systems.

We suggest that the major source of difference leading to the difference in complexity between the fully convective CTTS and our targets is the fact that our targets have developed a radiative core. This would then be analogous to the situation found in M dwarfs as recently reported (Donati et al. 2008c, Morin et al. 2008). The authors conduct spectropolarimetric surveys of M dwarfs ranging from  the higher mass stars with radiative cores down to  lower mass M stars that are close to  full convection. They report that the fully convective M dwarfs have simpler, large-scale almost fully poloidal fields compared to those found in early M dwarfs.
Reiners \& Basri (2009) have also found evidence for fully convective stars
having simpler fields using combined Stokes I and V data.
Clearly further observations of other pre-main sequence stars are needed to 
clarify whether or not the majority of CTTS with radiative cores show more 
complex fields than those in fully convective CTTS, but these initial
results would suggest that this is a strong possibility.

Reiners \& Basri (2009) point out that Stokes V studies alone may miss a significant fraction of the total magnetic flux in cool stars, even 
though Stokes V studies are still the most powerful diagnostics to probe the {\em geometry} of the stellar magnetic field. 
Future studies of T Tauri stars should ideally combine both Stokes I and Stokes V data, in order to measure the total 
magnetic flux in these systems as well as the field geometries. As these studies require contemporaneous optical and NIR spectra
of very variable systems, this is a challenging, yet potentially rewarding avenue for future studies.

\subsection{Evolutionary status and magnetic fields}

CV Cha and CR Cha are progenitors of early F and late A-type ZAMS stars, while V2129 Oph will become a F-type MS star. CV Cha and CR Cha should take under 17\,Myr to reach the ZAMS and less than 9\,Myr to become fully 
radiative  (Marconi \& Palla 2004). Over this time their magnetic fields should evolve rapidly to 
comply with the relatively weak  fields  observed in A-type stars, where only 10\% of stars are Ap stars and observably magnetic (Mathys 2004).  The two most likely explanations for the presence of magnetic fields in Ap-type stars are: slowly decaying fields that were originally frozen in from the ISM as the stars first formed,  or that the fields observed were originally generated in a pre-main sequence dynamo that ceased to operate once the star became fully radiative (Wade et al. 2008). Further ZDI observations of intermediate-mass PMS stars approaching the main sequence will be crucial in establishing whether or not similar magnetic fields exist in all intermediate-mass T Tauri stars and at what point this very strong ``cool star-type'' magnetic activity switches off during the approach to the main sequence.   

Evidently, there are also implications for the process of star formation in G and K-type stars too.  If future observations confirm our suggestion that 
stellar magnetic fields increase significantly in complexity as stars 
develop radiative cores, then this naturally alters the magnetosphere 
of the star and thus the efficiency with which the star can ``lock'' 
onto its surrounding disk.  This may be a significant factor in explaining 
how accretion finally stops in magnetically active stars. Detailed 
simulations of the effects of these changing fields are required to 
fully explore the effects of this significant evolution of the stellar 
magnetic field in magnetospheric accretion models. 

In modelling the magnetospheric accretion of V2129 Oph and BP Tau, 
Gregory et al. (2008)
conclude that the stronger dipole component of the fully-convective BP Tau may truncate the disk close to the co-rotation radius
in the manner assumed by many disk-locking models  (e.g.,  K\"onigl 1991, Collier Cameron 
\& Campbell, 1993). V2129 Oph, on the other hand, has a more complex field with a relatively weaker dipole
component which may allow the disk to penetrate closer to the star than the co-rotation radius. If the change
in the magnetic field structure of T Tauri stars as they develop radiative cores does indeed alter their ability to truncate their
accretion disks, it may also alter the balance of magnetic and accretion torques that governs the rotational evolution of
these stars.

This change in magnetic structure may also be apparent in the magnitude and rotational modulation of the X-ray  luminosity of these stars. 
For pre-main sequence stars in  NGC2264 ($\sim 3$\,Myr), Rebull et al. (2006)  found a significant drop in the X-ray luminosity of radiative 
compared to fully convective stars. If this is indeed related to the change in magnetic field structure, it may also be apparent in a greater 
rotational modulation in those stars with more complex fields. The smaller-scale fields of the stars with radiative cores may also be unable to 
support the very long field lines associated with the ``super-flares" detected in the COUP sample (Favata et al. 2005). These very powerful flares 
also exhibit the highest temperatures and produce the hardest X-ray photons, making them the favoured candidates for ionising the disk (Igea \& 
Glassgold 1999,  Ercolano et al 2008). The change in magnetic structure when the radiative core develops could therefore have a significant effect 
not only on the X-ray characteristics of the star, but also on the ionisation state and structure of the disk.

\begin{figure}
\includegraphics[width=9cm]{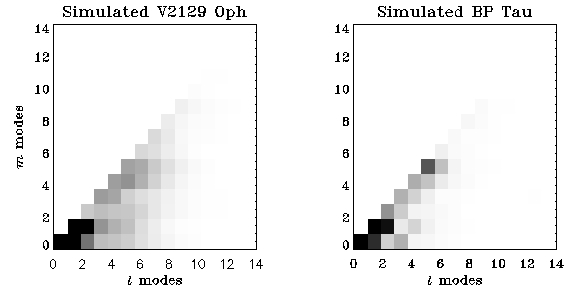}
\caption{Dominant $l,m$ modes for the simulated radial magnetic fields for V2129 Oph and BP Tau
  -- the input maps used were from CR Cha. Clearly the simulated and reconstructed ZDI V2129 Oph maps show  a similar level of complexity, while the simulated BP Tau
  map is significantly more complex than that obtained using ZDI (c.f. with Fig.\,\ref{fig:modes}).
}
\label{fig:simmodes}
\end{figure}

\appendix

\section{Spectral fits}
\begin{figure*}
\includegraphics[width=8cm,height=8.5cm]{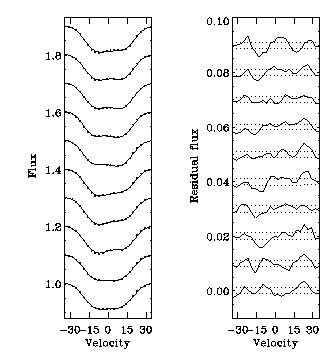}
\includegraphics[width=8cm,height=8.5cm]{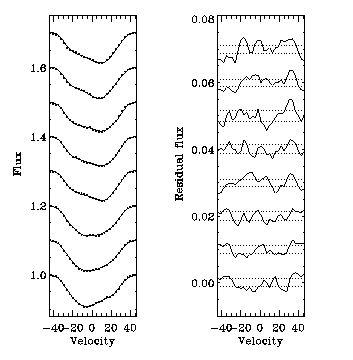}
\caption{Fits and residuals for the brightness maps (Fig.\,\ref{fig:intensity}) of CV Cha (left) and CR Cha (right).
Intensity maps for  both CV Cha and CR Cha produce fits to the dataset at a  
 \chisq$_r=2.5$. The profiles are stacked in order of observation as in Figs. 3 \& 4; see Table 2 for more details.}
\label{fig:intfits}
\end{figure*}

\begin{figure*}
\includegraphics[width=8cm,height=8.5cm ]{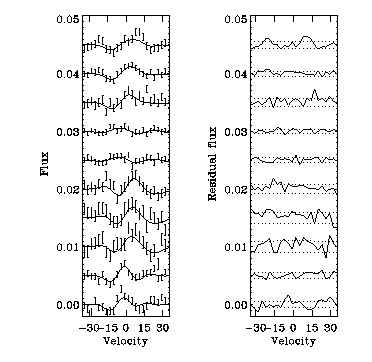}
\includegraphics[width=8cm,height=8.5cm]{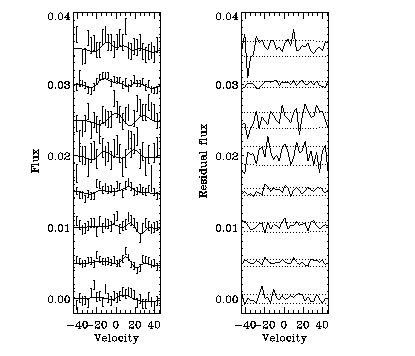}
\caption{Fits and residuals for the magnetic  field maps of CV Cha (left) and CR Cha (right).
The CV Cha map produces a fit of \chisq$=1.1$ and the CR Cha maps fit the observed spectra to a level of $\chi^2_r=1.0$. Residuals are shown on the right side of each plot with the $\pm 1$ error bar sizes shown as dashed lines. The profiles are shown stacked in order of observation as in Figs. 3 \& 4; see Table 2 for more details.
}
\label{fig:magfits}
\end{figure*}
\section*{Acknowledgments}
GAJH would like to thank Eike Guenther and Yvonne Unruh for useful discussions
that helped improve this paper. The observations in this paper were obtained at the Anglo-Australian Telescope and the assistance of the AAO staff with acquiring these observations is gratefully acknowledged. The authors also thank an anonymous referee for valuable suggestions that have improved this paper.

\label{lastpage}


\begin{thebibliography}{99}
\bibitem[\protect\citeauthoryear{Ardila \& Basri}{2000}]{ardila00} 
 Ardila, D.R., Basri, G., 2000, ApJ, 539, 834 
\bibitem[\protect\citeauthoryear{}{}]{} 
 Bally, J. Walawender, J., Luhman, K.L., Fazio, G., 2006, AJ 132, 1923
\bibitem[\protect\citeauthoryear{}{}]{}  
 Barnes, J.R., Collier Cameron, A., Unruh, Y.C., Donati, J.F., Hussain G.A.J., 1998, MNRAS, 299, 904
 \bibitem[\protect\citeauthoryear{}{}]{} 
Bary, J.S., Weintraub, D.A., Shukla, S.J., Leisenring, J.M., Kastner, J.H., 2008, ApJ 678, 1088
\bibitem[\protect\citeauthoryear{}{}]{} 
Bouvier, J., Bertout, C., Benz, W., Mayor, M., 1986, A\&A, 165, 110-119
\bibitem[\protect\citeauthoryear{}{}]{} 
Bouvier, J.,  Alencar,  S. H. P.,  Harries, T. J., Johns-Krull, C.M.,  Romanova, M.M., 2006, Protostars and Planets V, Tucson, Arizona, USA, Eds.  B. Reipurth, D. Jewitt,, K. Keil, p.479
\bibitem[\protect\citeauthoryear{}{}]{} 
Camenzind , M., 1990, Rev. Modern Astronomy, 3, 234
\bibitem[\protect\citeauthoryear{}{}]{} 
 Collier Cameron, A., Campbell, C.G., 1993, A\&A, 274, 309
\bibitem[\protect\citeauthoryear{}{}]{} 
Collier Cameron, A. \& Unruh, Y.C.,  1994, MNRAS, 269, 814
\bibitem[\protect\citeauthoryear{}{}]{} 
 Covino, E., Terranegra, L., Franchini, M., Chavvaria-K., C., Stalio, R., 1992, A\&A, 256, 525
\bibitem[\protect\citeauthoryear{}{}]{} 
 D'Antona, F. \& Mazzitelli, I., 1994, ApJS, 90, 467 
 \bibitem[Donati and Collier Cameron 1997]{donati97ab}
Donati, J.-F., Collier Cameron, A., 1997, MNRAS, 291, 1
\bibitem[Donati et al. 1997]{donati97lsd}
 Donati, J.-F., Semel, M., Carter, B.D., Rees, D.E., Collier Cameron, A., 1997, MNRAS, 291, 658
    \bibitem[\protect\citeauthoryear{}{}]{} 
Donati, J.-F., et al.,  2003, MNRAS, 345, 1145
\bibitem[\protect\citeauthoryear{}{}]{} 
Donati, J.-F.,  et al., 
2007, MNRAS, 380, 1297 
\bibitem[\protect\citeauthoryear{}{}]{} 
Donati, J.-F., et al., 
 2008a, 14th Cambridge Workshop on Cool Stars, Stellar Systems, and the Sun ASP Conf. Ser., Vol. 384, Pasadena, California, USA. Ed.  G van Belle., p.156
 \bibitem[\protect\citeauthoryear{}{}]{} 
Donati, J.-F., et al., 
2008b, MNRAS, 386, 1234
\bibitem[\protect\citeauthoryear{}{}]{} 
Donati, J.-F., et al., 
2008c, MNRAS, 390, 545
\bibitem[\protect\citeauthoryear{}{}]{} 
Ercolano, Barbara, Drake, Jeremy J., Raymond, John C., Clarke, Cathie C., 2008, ApJ, in press
\bibitem[\protect\citeauthoryear{}{}]{} 
Favata, F., Flaccomio, E., Reale, F., Micela, G., Sciortino, S., Shang, H., Stassun, K. G., Feigelson, E. D., 2005, ApJS, 160, 469
\bibitem[\protect\citeauthoryear{}{}]{} 
Feigelson, E.D., Casanova, S., Montmerle, T., Guibert, J., 1993, ApJ, 416, 623
\bibitem[\protect\citeauthoryear{}{}]{} 
Gauvin, L.S., Strom, K.M.,  1992, ApJ, 385, 217
\bibitem[\protect\citeauthoryear{}{}]{} 
Gregory, S. G., Matt, S. P., Donati, J. -F., Jardine, M., 2008, MNRAS, in press
\bibitem[\protect\citeauthoryear{}{}]{} 
Guenther, E.W.,  Esposito, M., Mundt, R., Covino, E., Alcal‡, J. M., Cusano, F., Stecklum, B.,  2007, A\&A, 467,  1147
\bibitem[\protect\citeauthoryear{}{}]{} 
 Herbst, W., Mundt, R., 2005, ApJ, 633, 967 
\bibitem[\protect\citeauthoryear{}{}]{} 
Herczeg, G.J., Hillenbrand, L.A., 2008, ApJ, 681, 594
\bibitem[\protect\citeauthoryear{}{}]{} 
Howarth, I.D., Murray J.,  Mills D. 1994, Starlink User Note, No. 50.15 
\bibitem[\protect\citeauthoryear{}{}]{} 
Hussain, G. A. J., Donati, J.-F., Collier Cameron, A., Barnes, J. R., 2000, MNRAS, 318, 961
\bibitem[\protect\citeauthoryear{}{}]{} 
Hussain, G.A.J., 2000, PhD thesis, Univ. St Andrews 
\bibitem[\protect\citeauthoryear{}{}]{} 
Igea, J., Glassgold, A. E., 1999, ApJ, 518, 848
\bibitem[\protect\citeauthoryear{}{}]{} 
 M. M. Jardine, S. G. Gregory, J.-F. Donati., 2008, MNRAS, 386, 688 
\bibitem[\protect\citeauthoryear{}{}]{} 
Johns-Krull, C., Valenti, J.A.,  Koresko, C., 1999, ApJ, 516, 900
\bibitem[\protect\citeauthoryear{}{}]{} 
Ko\"nigl, A., 1991, ApJ, 370, L39
\bibitem[\protect\citeauthoryear{}{}]{} 
Kupka F., Piskunov N.E., Ryabchikova T.A., Stempels H.C., Weiss W.W., 1999, A\&AS 138, 119-133. 
\bibitem[\protect\citeauthoryear{}{}]{} 
Kurucz, R. L., 1993, CDROM 13, 18 
\bibitem[\protect\citeauthoryear{}{}]{} 
Kurucz, R. L. 2005, MSAIS 8, 14 
\bibitem[\protect\citeauthoryear{}{}]{} 
Luhman, K. L. 2004, ApJ, 602, 816
\bibitem[\protect\citeauthoryear{}{}]{} 
Morin, J., et al., 
2008, MNRAS, 390, 567
\bibitem[\protect\citeauthoryear{}{}]{} 
 Marconi, M., Palla, F., 2004, ``The A-star puzzle'', Proc. IAU Symposium No. 224, Eds., J. Zverko, J. Ziznovsky, S.J. Adelman, W.W. Weiss 2004, p. 69
\bibitem[\protect\citeauthoryear{}{}]{} 
Mathys, G. 2004, in IAU Symp. 224, The A-Star Puzzle, eds.
J. Zverko, J. \v{Z}i\v{z}\v{n}ovsk\'y, S. J. Adelman,
\& W. W. Weiss (Cambridge Univ. Press), 225 
\bibitem[\protect\citeauthoryear{}{}]{} 
Natta, A. Testi, L., Muzerolle, J., Randich, S., Comeron, F., Persi, P., 2004, A\&A, 424, 603
\bibitem[\protect\citeauthoryear{}{}]{} 
Natta, A., Meyer, M.R., Beckwith, S.V. W., 2000, ApJ, 534, 838
\bibitem[\protect\citeauthoryear{}{}]{} 
Palla, F., Stahler, S.W., 1999, ApJ, 525, 882
\bibitem[\protect\citeauthoryear{}{}]{} 
Rebull, L. M., Stauffer, J. R., Ramirez, S. V., Flaccomio, E., Sciortino, S., Micela, G., Strom, S. E., Wolff, S. C., 2006, AJ, 131, 2934
\bibitem[\protect\citeauthoryear{}{}]{} 
Reiners, A., Basri, G., 2009, A\&A, in press
\bibitem[\protect\citeauthoryear{}{}]{} 
Reipurth, B., Pedrosa, A., Lago, M. T. V. T., 1996, A\&AS, 120, 229
\bibitem[\protect\citeauthoryear{}{}]{} 
Robrade, J., Schmitt, J.H.M.M., 2006, A\&A, 449, 737 
\bibitem[\protect\citeauthoryear{}{}]{} 
Schegerer, A., Wolf, S., Voshchinnikov, N.V., Przygodda, F., Kessler-Silacci, J.E., 2006, 
A\&A, 456, 535
\bibitem[\protect\citeauthoryear{}{}]{} 
Shevchenko, V.S., Herbst, W.,  1998, AJ, 116, 1419
\bibitem[\protect\citeauthoryear{}{}]{} 
Semel, M., 1989, A\&A, 225, 456  
\bibitem[\protect\citeauthoryear{}{}]{} 
Shu, F., Najita, J, Ostriker, E., Wilkin, F., Ruden, S., Lizano, S., 1994, ApJ, 429, 781
\bibitem[\protect\citeauthoryear{}{}]{} 
Siess, L., Dufour, E., Forestini, M., 2000, A\&A, 358, 593
\bibitem[\protect\citeauthoryear{}{}]{} 
Skelly, M.B., Unruh, Y.C., Collier Cameron, A., Barnes, J.R., Donati, J.-F., Lawson, W.A., Carter, B.D.,
2008, MNRAS, 385, 708
\bibitem[\protect\citeauthoryear{}{}]{} 
Stempels, H.C., Piskunov, N., 2003, A\&A, 408, 693 
\bibitem[\protect\citeauthoryear{}{}]{} 
Unruh, Y.C., Collier Cameron, A., Guenther, E.,  1998, MNRAS, 295, 781-798
\bibitem[\protect\citeauthoryear{}{}]{} 
Unruh, Y. C., et al., 
2004, MNRAS, 348, 1301-1320 
\bibitem[\protect\citeauthoryear{}{}]{} 
Valenti, J.A., Johns-Krull, C.M., Hatzes, A.P., 2003 in Brown, A., Harper, G., Ayres, T., eds. 12th Cambridge Workshop on Cool Stars Stellar Systems and the Sun, Univ. of Colorado, vol 12, p. 729
\bibitem[\protect\citeauthoryear{}{}]{} 
Wade, G.A., et al., 
2008, Proceedings of Solar Polarisation Workshop Number 5., ASP Conf. Series,   in press.
\bibitem[]{}
Wichmann, R., Bastian, U., Krautter, J., Jankovics, I., Rucinski, S. M., 1998, MNRAS, 301, 39L

\end{thebibliography}
\end{document}